\documentclass[11pt,article,twocolumn]{IEEEtran}
\IEEEoverridecommandlockouts

\usepackage{amsmath}
\usepackage{amssymb}
\usepackage{amsfonts}
\usepackage{float} 
\usepackage{srcltx}
\usepackage{mathrsfs} 
\usepackage{multirow}

\usepackage{graphicx}
\usepackage{epsfig}
\usepackage{psfrag}
\usepackage{subfigure}

\usepackage{setspace}
\usepackage{dsfont} 

\usepackage[]{units} 
\usepackage{url} 
\usepackage[dvips]{color}
\usepackage{verbatim} 
\usepackage[numbers,square,comma,compress]{natbib} 
\usepackage{ifthen} 
\usepackage{ifpdf} 

\usepackage{ulem} 
\usepackage{arydshln} 
\usepackage{stackrel} 
\usepackage{mathtools} 

\newcommand{\COMMENT}[1]{}
\newcommand{\rem}[1]{}

\newtheorem{example}{Example}

\begin{document}

\allowdisplaybreaks

\title{Comments on Downlink Non-Orthogonal Multiple Access:   Relative Gain Subject to Near Sum-Rate Optimality}

\author{Uri Erez}
\maketitle

\maketitle


\begin{abstract}
Non-orthogonal access techniques have recently gained renewed interest in the context of next generation wireless networks.
As the relative gain, with respect to traditionally employed orthogonal-access techniques depends on many factors, it is of interest to obtain  insights by considering the simplest scenario,
%
the two-user downlink (broadcast) channel where all nodes are equipped with a single antenna. Further, 
we focus on rate pairs that are in the vicinity of sum-rate optimalilty with respect to the capacity region of the broadcast channel.
A simple and explicit characterization of the relative gain of non-orthogonal transmission with respect to orthogonal transmission is obtained under these conditions as an immediate 
consequence of the capacity regions of the two. 


\end{abstract}

%
\section{Introduction}
\label{sec:intro}
Orthogonal multiple access (OMA) techniques have been at the core of 
cellular network communications to date. While constraining the 
users'  signals to be orthogonal in general comes at a price in terms 
of the achievable rate region, OMA techniques are nonetheless architecturally
appealing and are sum-rate optimal when all users are equipped with single antennas.

Despite the simplicity of OMA, non-orthogonal multiple access (NOMA) has gained increased attention over the last several years as a means of boosting communication rates in next generation cellular systems, 
 with a particular emphasis on increased fairness in the downlink between users with unbalanced signal-to-noise ratios (SNRs) 
 \cite{saito2013non}.
Indeed, the potential gains of NOMA over traditional OMA techniques have been extensively studied and reported in numerous recent works.

The relative gain, with respect to traditionally employed OMA techniques depends on many factors, including the relative SNRs of the different users, the utility function of the rate vector chosen, the power allocation policy, and the number of antennas used at both transmissions ends. 

It is therefore of interest to obtain simple insights on the relative gain of NOMA.
To that end, we restrict attention to the most basic scenario of (downlink) communication within a single cell where all nodes are equipped with a single antenna. We further focus attention on rate vectors that are in the vicinity of sum-rate optimality with respect to the information-theoretic capacity region. 

As the problem addressed has been widely studied, we make no claims of arriving at insights that have not been observed before. Rather, our goal is to capture the relative gains of downlink NOMA in a succinct  manner.

The downlink of a cellular communication link (with single-antenna nodes)
corresponds to a degraded broadcast channel, the information-theoretic limits of which are well understood \cite{CoverBook}, and are achieved by superposition coding. One may view power-domain NOMA on the downlink as synonymous with
information-theoretic (rate) optimal communication over the Gaussian broadcast channel. 
Thus,  the observations we make are but consequences of the well-established capacity region of this channel. 
The contribution of this note  is  in the  following  observations.

We define and quantify the relative gain of  NOMA (with respect to OMA) as the ratio of the rates achieved for the weaker user by the two respective schemes, for a given (common) rate allocated to the stronger user. More specifically, we give a simple expression that bounds this ratio that is tight when the former rates are small, i.e., in the vicinity of sum-rate optimality. 

It is observed that while sum-rate optimality requires serving only the strongest user,  if both users are at high SNR conditions, then one may shift away some rate from the stronger user to the weaker one with little loss (measured as a fraction of the rate shifted) to the sum-rate, to a first order approximation (in the shifted rate).

Finally, the analysis corroborates and  quantifies the assertion that for a given SNR of the strong user, the relative gain of NOMA is greatest when the weaker user is at low SNR. 



\section{Relative Gain}
\label{sec:downlink}
For simplicity, we consider a downlink channel 
with only two users. The signal received at user 
$i$ can be written as
\begin{align}
y_i=h_i x +n_i, 
\end{align}
where the transmitted signal  $x$ is subject to 
the power constraint $P$, $n_i$ is circularly-symmetric complex Gaussian noise with power $N$, and $h_i$ denotes the channel coefficient. 
We define $S_i=\frac{P|h_i|^2}{N}$ as the SNR of user $i$
where without loss of generality we assume that 
$S_1 \geq S_2$. 
We further assume that the SNRs are known to the transmitter. 

It is well known that for balanced SNRs, i.e.,
when $S_1=S_2$, OMA achieves the capacity region of the channel. It is therefore  clear that in order for NOMA to have a substantial gain over OMA, the ratio $S_1/S_2$ needs to be sufficiently large. In the analysis 
to follow we quantify this statement and also study how the gain depends on the SNR pair (beyond the ratio of the two).

The downlink channel is a degraded Gaussian broadcast channel for which the capacity region is well known. Every rate point in the capacity region may be achieved by superposition coding (power-domain NOMA) and successive decoding \cite{CoverBook,tse2005fundamentals}.
Namely, let us denote the by $0 \leq a \leq 1$ the 
fraction of the total power allocated to user 1, 
so that the transmitted signal can be written as
\begin{align}
x=\sqrt{a} \, x_1+\sqrt{1-a} \, x_2.
\end{align}
Then the capacity region is given by all rate pairs
$(R_1,R_2)$ satisfying 
\begin{align}
&R_1  \leq  \log (1+a S_1)   \nonumber  \\
&R_2  \leq \log \left(1+\frac{(1-a) S_2}{1+a S_2} \right),  
\label{regionNOMA}
\end{align}
for some $0 \leq a \leq 1$.

In contrast, the achievable rate region of OMA transmission is 
given by all rate pairs satisfying 
\begin{align}
&R_1  \leq  \alpha \log (1+S_1)   \nonumber  \\
&R_2  \leq (1-\alpha) \log \left(1+S_2 \right),
\label{regionOMA}
\end{align}
from some $0 \leq \alpha \leq 1$, where $\alpha$ is a time-sharing parameter.

\begin{figure}
    \centering
    \includegraphics[scale=.45]{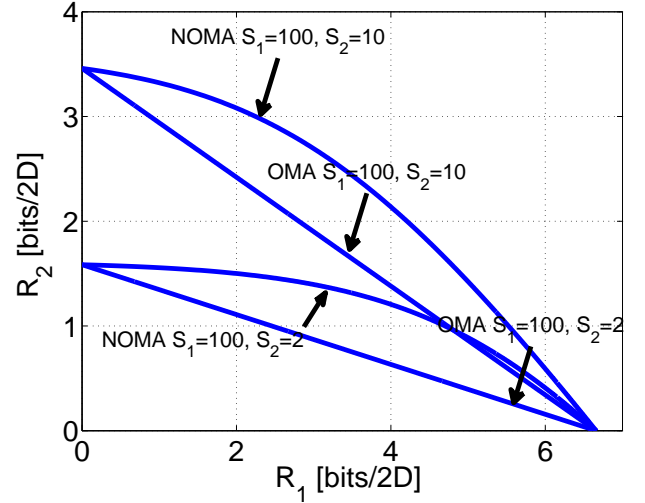}
    \caption{Downlink: Achievable rate regions of NOMA and OMA for two pairs of SNRs: $S_1=100,S_2=10$ and $S_1=100,S_2=2$. }
    \label{fig:BC}
\end{figure}

The capacity region is depicted in Figure~\ref{fig:BC} for 
two different pairs of SNRs. Clearly, to maximize  throughput, one would allocate all the resources to the stronger user, in which case OMA and NOMA trivially coincide. While such a rate allocation is invalid under any reasonable utility function (leaving scheduling aside), we argue that one may nonetheless gain considerable insight by perturbing off of this rate-pair point.  To that end, let us define  (the magnitude of) the slope at the 
sum-rate optimal point:
\begin{align}
{\rm slope}=- \left. \frac{d R_2}{d R_1} \right|_{R_1=\log(1+S_1)},
\end{align}
where the slope  is understood to be calculated according to either (\ref{regionNOMA}) or (\ref{regionOMA}), in correspondence to the scheme considered.  
Thus,
\begin{align}
{\rm slope}_{\rm OMA}=
\frac{\log(1+S_2)}{\log(1+S_1)}.
\label{slopeOMA}
\end{align}
For NOMA, a straightforward calculation yields
\begin{align}
{\rm slope}_{\rm NOMA}=
\frac{S_2}{S_1}
\cdot
\frac{1+S_1}{1+S_2}.
\label{slopeNOMA}
\end{align}
We refer to the ratio 
\begin{align}
g(S_1,S_2)={\rm slope}_{\rm NOMA}/{\rm slope}_{\rm OMA}
\label{rel_gain}
\end{align}
as the relative gain of NOMA. The defined gain should be understood as the maximal possible gain since as we move further away from the sum-rate optimal point (i.e., as $R_1$ decreases), the actual gain (ratio of rates achieved by the weaker user) will decrease.

\begin{figure}
    \centering
    \includegraphics[scale=.45]{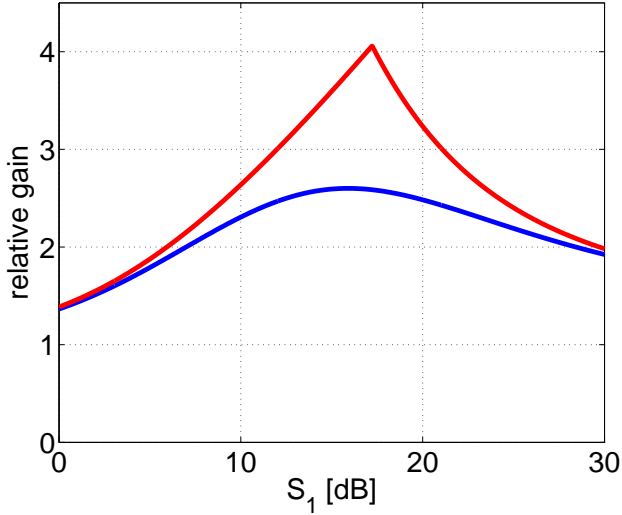}
    \caption{Relative gain: The lower curve is the relative gain as a function of $S_1$, where $S_2$ is $15$dB smaller than $S_1$. The upper curve is the minimum of the two bounds (\ref{rel_gain_bound}) and (\ref{rel_gain_bound2}).}
    \label{fig:gain}
\end{figure}

Two simple upper bounds on the relative gain are easily obtained.
Using the well-known inequality 
$\log(1+x)\geq \log(e) \cdot \frac{x}{1+x}$, 
we obtain:
\begin{align}
g(S_1,S_2)\leq \frac{\log(1+S_1)}{\log(e)} \cdot \frac{1+S_1}{S_1}.
\label{rel_gain_bound}
\end{align}
Also, since ${\rm slope}_{\rm NOMA}\leq1$, it follows that,
\begin{align}
g(S_1,S_2)   \leq  \frac{1}{{\rm slope}_{\rm OMA}} = \frac{\log(1+S_1)}{\log(1+S_2)}.
\label{rel_gain_bound2}
\end{align}
Figure~\ref{fig:gain} depicts an example of the relative gain along with the bounds. As can be seen, and will become evident in the sequel, for a fixed SNR gap (in dB), the relative gain as a function of $S_1$ starts at $1$ for $S_1 \ll 1$ and ends at $1$ for $S_1 \gg  1$, and hence the maximal gain lies somewhere in between. 

In order to attain further insight into the relative gain of
NOMA, we next consider separately the different possible SNR regimes.
\subsection{Low SNR regime: no gain}
Due to the fact that $\log(1+x) \approx x \log(e) $ for $x \ll 1$, when both
users are at (very) low SNR, the capacity regions of OMA and NOMA coincide and the relative gain approaches $1$. 
\subsection{Mixed SNR regime: maximal gain}
When $S_1 \gg 1$ (and general value of $S_2$),
(\ref{slopeNOMA}) becomes
\begin{align}
{\rm slope}_{\rm NOMA}\approx
\frac{S_2}{1+S_2}.
\label{slopeNOMA2}
\end{align}
This is quite different from the behavior of OMA where the slope can be arbitrarily small as $S_1$ grows while $S_2$ is held fixed. 
Indeed, for $S_2 \ll 1$, we obtain
\begin{align}
g(S_1,S_2)\approx
\log(S_1)/\log(e).
\label{gain_mixed}
\end{align}
This is (approximately) the maximal relative gain that can be attained by NOMA for large $S_1$. 
More precisely, the bound (\ref{rel_gain_bound}) is tight when $S_2 \ll 1$ and reduces to (\ref{gain_mixed}) when in addition $S_1 \gg 1$.

The relative gain can be quite substantial at typical SNR ranges when pairing a weak and strong user. 
We note, however, that while indeed the weak user may well be at low SNR, the condition $S_2 \ll 1$ will likely not hold. Thus,  one should view (\ref{gain_mixed}) as an (under normal circumstances, unattainable) upper bound on the relative gain of NOMA. 
A reasonable ``rule of thumb" for the maximal gain in a typical setting 
can be obtained by assuming large $S_1$ and setting $S_2=1$ in (\ref{slopeOMA}), (\ref{slopeNOMA}), and (\ref{rel_gain}), yielding
$g(S_1,S_2) \sim \nicefrac{1}{2}\log_2(S_1)$.

\begin{example}
As an example, suppose that $S_1=20$ ($13 $ dB) and $S_2=1$ ($0$ dB). Then, by computing 
(\ref{slopeOMA}) and
(\ref{slopeNOMA})  we obtain  ${\rm slope}_{\rm NOMA}=0.53$ and ${\rm slope}_{\rm OMA}=0.23$ so the 
relative gain (their ratio) amounts to $2.31$ whereas the mixed regime approximation as given in (\ref{gain_mixed}) amounts to a gain of $3$. The maximal sum rate is attained by setting  $a=1$ for both NOMA and OMA, which yields the rate pair $R_1=4.3923$ and $R_2=0$. Now suppose we decrease the rate of user $1$ by one bit. 
This is achieved for NOMA by setting $a=0.475$ in (\ref{regionNOMA}), giving the weaker user a rate of   $R_2=0.44$. 
For OMA, a rate reduction of one bit for user $1$ is achieved by setting 
$a=0.77$ in (\ref{regionOMA}) which 
gives $R_2=0.23$. The relative gain is thus 
$1.93$ which is reasonably close to the limiting gain of $2.31$ predicted by 
(\ref{rel_gain}). 
\end{example}

As we see next, the weaker user need not be at low SNR for NOMA to have a significant relative gain.

\subsection{High SNR regime: some gain}
\label{sec:high}
When both $S_1 \gg 1$ and $S_2 \gg 1$,
(\ref{slopeNOMA}) becomes \mbox{${\rm slope}_{\rm NOMA}\approx 1$} and therefore the bound (\ref{rel_gain_bound}) becomes tight, and the relative gain reduces to
\begin{align}
g(S_1,S_2)\approx
\frac{\log(S_1)}{\log(S_2)}.
\end{align}
While not as large as in the mixed SNR regime, the relative gain can still be substantial for large SNR differences. 

It is interesting to note that for NOMA in the high SNR regime, since the slope is one, the sum rate remains unchanged when we start shifting rate from the stronger user to the weaker user. In other words, at high SNR, fairness comes for free to a first order approximation in the weaker user's rate (around zero).

\bibliographystyle{IEEEtran}


\end{document}